\documentclass[conference]{IEEEtran}
\IEEEoverridecommandlockouts
\usepackage{cite}
\usepackage{amsmath,amssymb,amsfonts}
\usepackage{algorithmic}
\usepackage{graphicx}
\usepackage{textcomp}
\usepackage{xcolor}
\usepackage{color}
\usepackage{balance}
\def\BibTeX{{\rm B\kern-.05em{\sc i\kern-.025em b}\kern-.08em
    T\kern-.1667em\lower.7ex\hbox{E}\kern-.125emX}}

\begin{document}

\title{ \Large A Secure and Smart Framework for Preventing Ransomware Attack}

\author{\IEEEauthorblockN{Jaspreet Kaur }
\IEEEauthorblockA{\textit{PhD Scholar at CSE Department}
 \\
\textit{Indian Institute of Technology Jodhpur}\\
Jodhpur, India \\
kaur.3@iitj.ac.in}}

\maketitle
 \begin{abstract}
Nowadays security is major concern for any user connected to the internet. Various types of attacks are to be performed by intruders to obtaining user information as man-in-middle attack, denial of service, malware attacks etc. Malware attacks specifically ransomware attack become very famous recently. Ransomware attack threaten the users by encrypting their most valuable data, lock the user screen, play some random videos and by various more means. Finally attacker take benefits by users through paid ransom. In this paper, we propose a framework which prevent the ransomware attack more appropriately using various techniques as blockchain, honeypot, cloud \& edge computing. This framework is analysed mainly through the IoT devices and generalized to the any malware attack.    
\end{abstract}


\section{Introduction}
There are lots of malware attack happen today but ransomware is one of the most dangerous one that threaten its
victim. There are various types of ransomware as Lock-type ransomware, Crypto-ransomware etc. It usually install by as some another malware
through malicious e-mail attachments or links, infected software
applications or websites and via some connected external storage systems. Then check the vulnerabilities of the system by which make a connection to the C \& C server and encrypt users important data or lock the screen of the users etc. Finally attackers demand the payment as cryptocurrency or threaten the users till they are not satisfied and delete or hide their traces so that no one can easily track them.\\\\
There are various approaches mentioned in literature for detecting and preventing this attack as signature based and anomaly based techniques which takes various parameters as input- open ports, various system calls, registry editing, accessing of files and folders, file entropy changes etc. and make prediction of attack. Some of them use sandboxing methods and use cloud storage for backup purpose. But all of them have some weaknesses as need large storage for signatures, not predicting zero day attacks, high false alarm rate, new advanced ransomware evading from sandbox by sleep mode at earlier time, only for the user mode but what if- when root is compromised, training database or machine learning or deep learning algorithms are compromised.
These above limitations motivate us to develop a new framework which can efficiently detect and mitigate this attack.
\section{Proposed Framework}
In our proposed method (figure 1), we take a example of a smart home environment in which various smart heterogeneous (both resource constrained and high resource power) IoT devices are to be vulnerable from this attack. We create a database using blockchain smart contract in which authentication, access control of IoT devices are to be stored for removing the risk of external device attack. This smart contract also contains the valuable system call functions as registry editing functions,  root access functions, cryptographic encryption functions, port assessing functions, desktop controlling functions, file read or write functions, folder deleting functions etc. These calls are included in the basic steps of ransomware attack, when intruder performs  sequence of operations as fingerprinting, port scanning, connection to the C \& C server, accessing of important data, encrypting the important documents, lock the user screen, demand  for the payment and delete the backup data along with traces of intruders. Some of the calls same as the normal operations and others are not but it makes difference when executed in sequence of ransomware attack.\\\\
But due to the complex nature of blockchain operations (mining and storage), it cannot be stored at every IoT devices specifically resource constrained devices. That's why we use edge computing here, means take  high power IoT devices at edges which compute all blockchain operations and data analysis. For security reason key management and generation are to be done at particular device itself.
\begin{figure*}[h]
\centering
    \includegraphics[width=16cm,height=7.5cm]{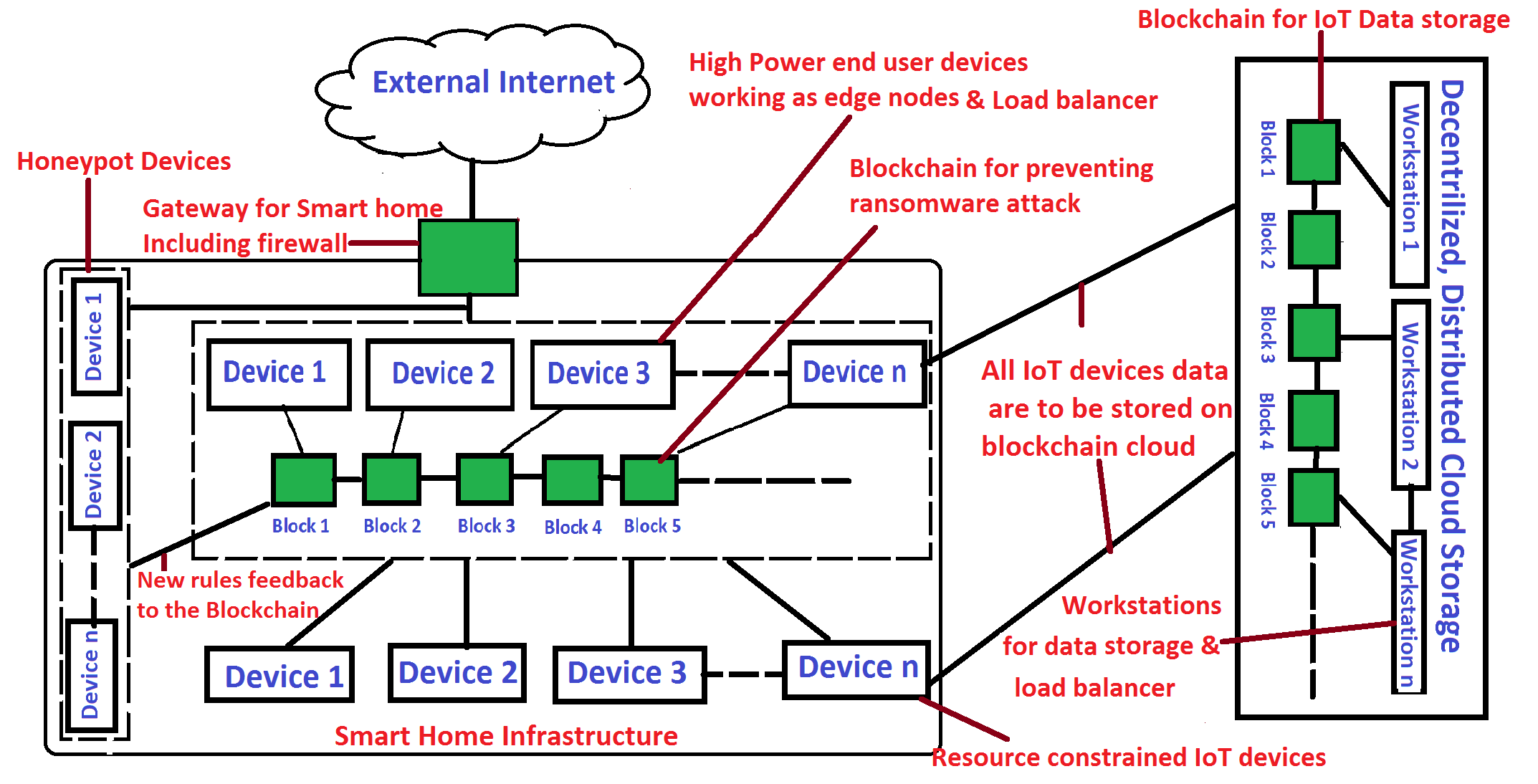}
    \caption{Proposed Approach- A Secure and Smart Framework for Preventing Ransomware Attack}
    \label{fig:my_label}
\end{figure*}
We also take some honeypot devices (very high power devices) for proper evaluation of this attack before that attack is performed on the actual device. For backup data of these IoT devices, we use cloud architecture which is also decentralized, distributed in nature mean managed by blockchain. Both of the blockchain managed by the load balancer devices or software for handling incoming request messages. For additional security,
the smart home gateway includes the firewall for filtering malicious ports, IP addresses, file extensions etc. to preventing ransomware attack. 

\section{Working Process}
 Any ransomware or software come to the smart home infrastructure either via network or through physical devices. For network attack, it first go through the gateway that include first line of defence as firewall which filter malicious ports, IP addresses, file extensions etc. But advanced malware can easily by pass it. Following steps are same for both (device and network attack). This software or malware run at 2 environment simultaneously-first one at that device which they want to access and second one on  the honeypot devices. Blockchain smart contract check the various sequences of system call run by that software and also check the authenticity and access control of that external device. If that software check for searching of a file, read/write of files, need admin privilege, edit log and registry events etc. Then this software behaviour is same as normal user behaviour. But when it wants to access network information as open ports, open sessions, read the user's browsing
history and bookmarks and read kernel level information then that smart contract alert the user because it may be the malware and require the user permission for execute it. After some time, if that software  call any encryption function, call to make a new network connection using tor browser, call to disable key guard and need the control over the desktop, then our blockchain stop this software processing at all devices except the honeypot nodes before that functions are executed and wait till software execution end up at that honeypot devices, if that software demand the ransom then it is a ransomware attack. Uninstall that software at all the devices and check honeypot devices for some new rules given as feedback to the addition of blockchain rules for more efficient detection of attacks.
For secure backup data of these IoT devices, we use blockchain cloud infrastructure with load balancer (schedulers) and various workstations. All blockchain operations  are to be maintained at cloud but key management are done at devices themselves for providing end point security. 

\section{Theoretical Analysis}
 We perform the theoretical analysis for validation of our work. Some of the observations are as:
 \begin{enumerate}
  \setlength{\itemsep}{0pt}
  \setlength{\parskip}{0pt}
     \item By the use of ransomware blockchain, attacker cannot compromised our data analysis algorithms or methods.
     \item Zero day attacks are prevented by the honeypot devices and new attacks rules are to be added at blockchain for more efficient detection of attacks.
     \item All IoT devices authenticates by the blockchain so that no external device attack in the network.
     \item Basic firewall rules are applied at gateway for novice intruder.
     \item For backup data securely, cloud blockchain is used.
     \item For end point security, key management are to be done at devices themselves for both of the blockchain.
     \item For complex operations of blockchain, edge and cloud infrastructure are to be used along with load balancer (hardware or software) for handling incoming requests.  
 \end{enumerate}
\section{CONCLUSION AND FUTURE WORK}
Nowadays ransomware attack is one of the most dangerous attack. It can very easily spread to any vulnerable system or device and demand the ransom from users. Various techniques are available in literature for mitigating this attack but every solution has problem as handling zero day attack, high false alarm rates, security of machine learning data and methods etc. So, for overcoming to these limitations we propose a new smart and secure method which includes blockchain, honeypot, edge or cloud computing techniques to IoT devices for preventing this attack. Our method is generalized means can be extended to any malware attack detection. In future work, firstly we implement our methodology on real world scenario and do changes in it as needed.

\end{document}